\documentclass{PoS}
\usepackage{bbm, float, setspace, epsfig, wrapfig, picins}
\usepackage{amsmath, amsfonts, subfigure, amssymb, fontenc, times, mathptmx, graphicx}
\pdfoutput=1

\title{Monte Carlo study of a Fermi gas with infinite scattering length}

\ShortTitle{Monte Carlo study of a Fermi gas with infinite scattering length}

\author{\speaker{O. Goulko}\\
        DAMTP, University of Cambridge, Wilberforce Road, Cambridge CB3 0WA, UK\\
        E-mail: \email{O.Goulko@damtp.cam.ac.uk}}

\author{M. Wingate\\
        DAMTP, University of Cambridge, Wilberforce Road, Cambridge CB3 0WA, UK\\
        E-mail: \email{M.Wingate@damtp.cam.ac.uk}}

\abstract{The Fermi gas at unitarity is a particularly interesting system of cold atoms, being dilute and strongly interacting at the same time. It can be studied non-perturbatively with Monte Carlo methods, like the recently developed worm algorithm. We discuss our implementation and tests of this algorithm and suggest a modification that increases its efficiency by reducing autocorrelations. We then show how the worm algorithm can be applied to calculate the critical temperature of an imbalanced Fermi gas (unequal number of fermions in the two spin components). We finally present some results obtained with the modified algorithm, in the balanced as well as in the imbalanced case.}

\FullConference{The XXVII International Symposium on Lattice Field Theory\\
                 July 26-31, 2009\\
                 Peking University, Beijing, China}

\begin{document}

\renewcommand{\Im}{\textnormal{Im}}
\newcommand{\sgn}{\textnormal{sign}}

\section{Introduction}
Lattice methods are useful for studying strongly interacting theories in particle as well as in condensed matter physics. When strong interactions render a perturbative study impossible, lattice field theory can provide a useful tool for numerical calculations. The Fermi gas at unitarity is a prominent example for such a strongly interacting system \cite{review}.

Fermionic matter is ubiquitous in nature, from the electrons in metals and semiconductors or the neutrons in the inner crust of neutron stars, to gases of fermionic atoms, like $^{40}$K or $^6$Li that can be created and studied under laboratory conditions. Due to Fermi-Dirac statistics, a dilute system of spin-polarised fermions exhibits no interactions and can be viewed as an ideal Fermi gas. However, interactions become crucial when we are dealing with fermions of several spin species. Low-energy interactions are characterised by the scattering length $a$. An especially intriguing case is the Fermi gas at divergent scattering length -- the unitary regime, in which the gas is dilute (range of potential $\ll$ interparticle distance) and strongly interacting (interparticle distance $\ll$ scattering length) at the same time. One key feature of this regime is universality: since all information about interactions is contained in the scattering length, and this length scale is no longer present at unitarity, the gas exhibits universal behaviour that only depends on two dimensionful parameters, temperature and density.

Due to a lack of an exact theoretical description several approximate and numerical approaches have been tried to study the Fermi gas in the unitarity limit. An accurate result for the critical temperature has been obtained with the recently developed Diagrammatic Determinant Monte Carlo (DDMC) algorithm \cite{main}. In the following we will first introduce the model and the algorithm and then present our modifications and results.

\section{Fermi-Hubbard model at finite temperature}

The Fermi-Hubbard model is the simplest lattice model for two-particle scattering. Its Hamiltonian in the grand canonical ensemble is given by
\begin{equation}
H=H_0+H_1=\sum_{\mathbf{k},\sigma}(\epsilon_\mathbf{k}-\mu_\sigma)c^\dagger_{\mathbf{k}\sigma}c_{\mathbf{k}\sigma}+U\sum_{\mathbf{x}}c^\dagger_{\mathbf{x}\uparrow}c_{\mathbf{x}\uparrow}c^\dagger_{\mathbf{x}\downarrow}c_{\mathbf{x}\downarrow},
\end{equation}
where $\epsilon_\mathbf{k}=\frac{1}{m}\sum_{j=1}^{3}(1-\cos{k_j})$ is the discrete dispersion relation, and $c^\dagger_{\mathbf{k}\sigma}$ ($c_{\mathbf{k}\sigma}$) the time-dependent fermionic creation (annihilation) operator. The model assumes non-relativistic fermions of two species labelled by $\sigma$ (which we will call "spin up" and "spin down") with equal particle mass $m$. For the present we also assume equal chemical potential $\mu_\uparrow=\mu_\downarrow\equiv\mu$ for the two spin species. The attractive contact interaction is characterised by the coupling constant $U<0$. The limit of infinite scattering length corresponds to $U=-7.914$, in units where $m=1/2$. We work on a 3D simple cubic spatial lattice with $L^3$ sites, periodic boundary conditions and lattice spacing set to unity. The time direction remains continuous. The continuum limit of this model can be taken by extrapolation to vanishing filling factor.

\begin{figure}[t]
\includegraphics[width=\textwidth]{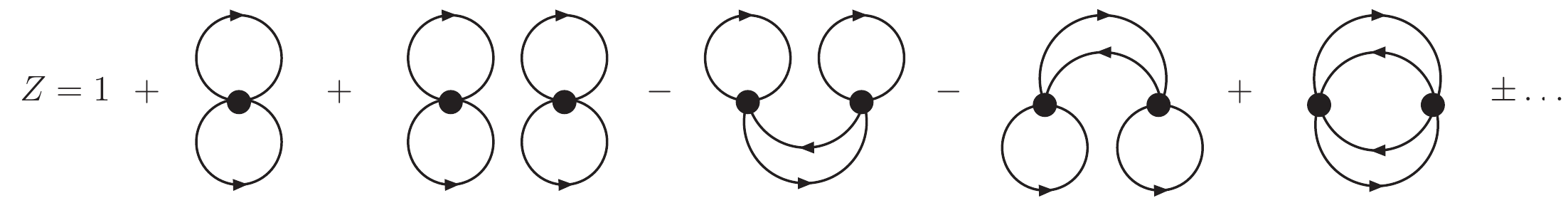}
\caption{Diagrammatic expansion of the partition function}
\label{Zexp}
\end{figure}

To study the finite temperature behaviour consider the grand canonical partition function in the imaginary time interaction picture,
\begin{equation}
Z=\textnormal{Tr}e^{-\beta H}=\textnormal{Tr}e^{-\beta H_0}\mathbf{T}_\tau \exp\left(-\int_0^\beta d\tau H_1(\tau)\right),
\end{equation}
where $\beta$ is the inverse temperature, $\mathbf{T}_\tau$ the imaginary time ordering operator and $H_1(\tau)=e^{\tau H_0}H_1e^{-\tau H_0}$. Expanding $Z$ in powers of $H_1$,
\begin{equation}
Z=\sum_{p=0}^{\infty}(-U)^p\sum_{\mathbf{x}_1,\ldots\mathbf{x}_p}\int_{0<\tau_1<\ldots\beta}\prod_{j=1}^{p}d\tau_j\textnormal{Tr}\left[e^{-\beta H_0}\prod_{j=1}^{p}c_\uparrow^\dagger(\mathbf{x}_j,\tau_j)c_\uparrow(\mathbf{x}_j,\tau_j)c_\downarrow^\dagger(\mathbf{x}_j,\tau_j)c_\downarrow(\mathbf{x}_j,\tau_j)\right],
\end{equation}
generates a series of Feynman diagrams, as shown in figure \ref{Zexp}.
Since we are ultimately interested in thermal expectation values of operators and thermal averages are calculated using the expansion of the partition function, it would be convenient to use this expansion as a probability distribution to generate configurations for Monte Carlo sampling. However, each fermionic loop contributes a minus sign, with the consequence that the diagrams in the series have different signs. For our purpose we need to rewrite the series as a sum of positive terms only. This can be done by considering all diagrams of order $p$ as one entity, which can be represented as the product of two matrix determinants (one $p\times p$ matrix for each spin component). The partition function can then be written as
\begin{equation}
Z=\sum_{S_p}(-U)^p\det\mathbf{A}^{\uparrow}(S_p)\det\mathbf{A}^{\downarrow}(S_p)
\label{partitionfunction}
\end{equation}
where $S_p$ denotes a vertex configuration (the spacetime positions of all vertices) and the matrix entries are free finite-temperature single-particle propagators $A^\sigma_{ij}(S_p)=G_{(0)}^\sigma(\mathbf{x}_i-\mathbf{x}_j,\tau_i-\tau_j)$. If the chemical potential is equal for spin up and spin down fermions (the balanced case) we have $\det \mathbf{A}^\uparrow\det\mathbf{A}^\downarrow=|\det\mathbf{A}|^2$, so that all terms in the series are positive \cite{rubtsov}.

The physical observable in the focus of our study is the order parameter for the phase transition to superfluidity. We introduce the pair creation and annihilation operators $P(\mathbf{x},\tau)=c_{\mathbf{x}\uparrow}(\tau)c_{\mathbf{x}\downarrow}(\tau)$ and $P^\dagger(\mathbf{x}',\tau')=c^\dagger_{\mathbf{x}'\uparrow}(\tau')c^\dagger_{\mathbf{x}'\downarrow}(\tau')$. At the critical point the correlation function
\begin{equation}
G_2(\mathbf{x}\tau;\mathbf{x}'\tau')=\left\langle\mathbf{T}_\tau P(\mathbf{x},\tau)P^\dagger(\mathbf{x}',\tau')\right\rangle=\frac{1}{Z}\textnormal{Tr}[\mathbf{T}_\tau P(\mathbf{x},\tau)P^\dagger(\mathbf{x}',\tau')e^{-\beta H}]
\end{equation}
is proportional to $|\mathbf{x}-\mathbf{x}'|^{-(1+\eta)}$ as $|\mathbf{x}-\mathbf{x}'|\rightarrow\infty$, where $\eta\approx0.038$ is the anomalous dimension for the U$(1)$ universality class. The integrated correlation function
\begin{equation}
R(L,T)\equiv L^{1+\eta}(\beta L^3)^{-2}\sum_{\mathbf{x},\mathbf{x}'}\int_{0}^{\beta}d\tau\int_{0}^{\beta}d\tau'G_2(\mathbf{x}\tau;\mathbf{x}'\tau')
\label{intcorrfn}
\end{equation}
will be independent of the lattice size $L$ at $\beta_c=1/T_c$. This property can be used to determine the critical temperature.

\section{Worm algorithm}

The configuration space is sampled via a Monte Carlo Markov chain process: in each step one of the possible updates to another vertex configuration is proposed, and accepted with some probability given by the detailed balance equations. The requirements of detailed balance and ergodicity ensure that the produced configurations are indeed distributed according to the correct thermal probability distribution $\rho(S_p)=\frac{1}{Z}(-U)^p|\det\mathbf{A}(S_p)|^2$, given by the expansion of the partition function.

The diagrammatic expansion of $\textnormal{Tr}[\mathbf{T}_\tau P(\mathbf{x},\tau)P^\dagger(\mathbf{x}',\tau')e^{-\beta H}]$ is similar to that of $Z$, but contains an additional pair of 2-point vertices at $(\mathbf{x},\tau)$ and $(\mathbf{x}',\tau')$. It is thus of advantage to sample these two series in the same simulation. In addition to sampling the regular 4-point diagrams we allow updates that insert the pair of 2-point vertices ("worm vertices") into the configuration space. This setup has several advantages. Firstly, the Monte Carlo estimator for the order parameter becomes very simple: it is the ratio of configurations with and without worm vertices. Secondly, all updates can now be performed through the worm vertices $P$ and $P^\dagger$, which simplifies the setup.

A detailed description of the individual updates can be found in the appendix of \cite{main}. Here we will only give a brief summary.

\begin{itemize}
\item Updates only concerning the worm vertices:
\begin{itemize}
\item Worm creation/annihilation: insert/remove the pair $P(\mathbf{x},\tau),\ P^\dagger(\mathbf{x}',\tau')$ into/from the configuration.
\item Worm shift: shift the $P^\dagger(\mathbf{x}',\tau')$ vertex to other coordinates.
\end{itemize}
\item Updates of the regular 4-point vertices: adding/removing a 4-point vertex (changes the diagram order).
\begin{itemize}
\item Diagonal version: add or remove a random vertex.
\item Alternative using worm: move the $P(\mathbf{x},\tau)$ vertex to another position and insert a 4-point vertex at its old position. The new coordinates are chosen in a way that tends to prolong existing vertex chains. In this case an update can only happen when the pair of 2-point vertices is present.
\end{itemize}
\end{itemize}

The worm setup, as proposed in \cite{main}, leads to much higher acceptance rates than the regular "diagonal setup". The idea behind the worm setup is that at low densities the major contribution comes from multi-ladder diagrams, these are configurations where the vertices are arranged into several vertex chains. The typical size of a chain depends on the parameters of the system. To favour the creation of vertex chains the worm update uses the 2-point vertex $P$ to add (or remove) 4-point vertices in a small spacetime region, while $P$ gets shifted to new coordinates each time. At low densities the new coordinates of $P$ will be chosen according to a probability distribution that favours zero or small spatial shifts and small temporal shifts in the direction of positive $\tau$. The removal update always attempts to remove the nearest neighbour of $P$, which means that due to detailed balance the addition update can only be accepted if the added vertex is the nearest neighbour of the shifted $P$-vertex. This nearest neighbour condition is crucial for achieving high acceptance rates.

\section{Modifications of the algorithm}

\begin{figure}[h!]
\begin{center}
\subfigure[Diagonal setup]{\label{endiag}\includegraphics[width=.49\textwidth]{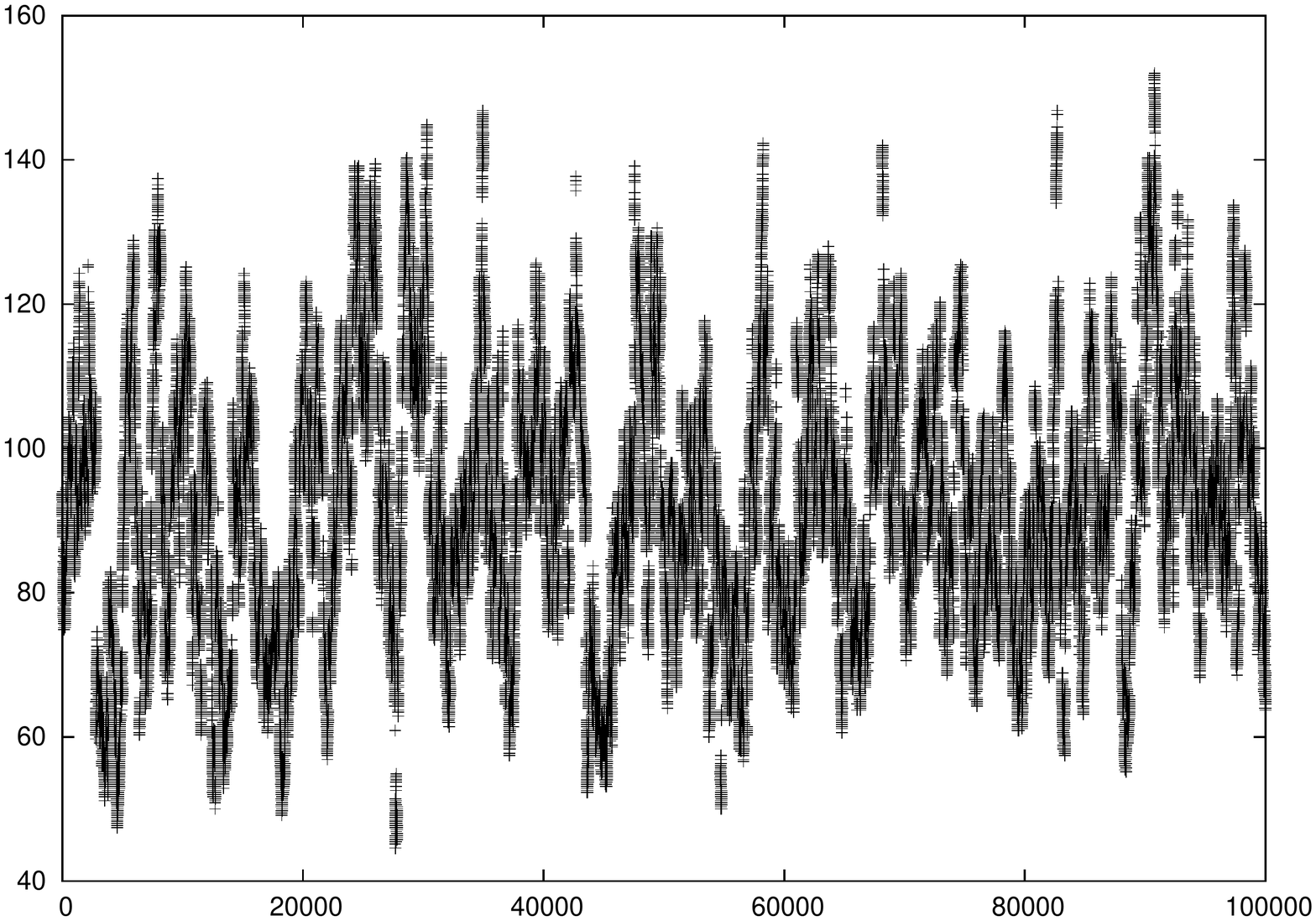}}
\hfill
\subfigure[Worm setup]{\label{enworm}\includegraphics[width=.49\textwidth]{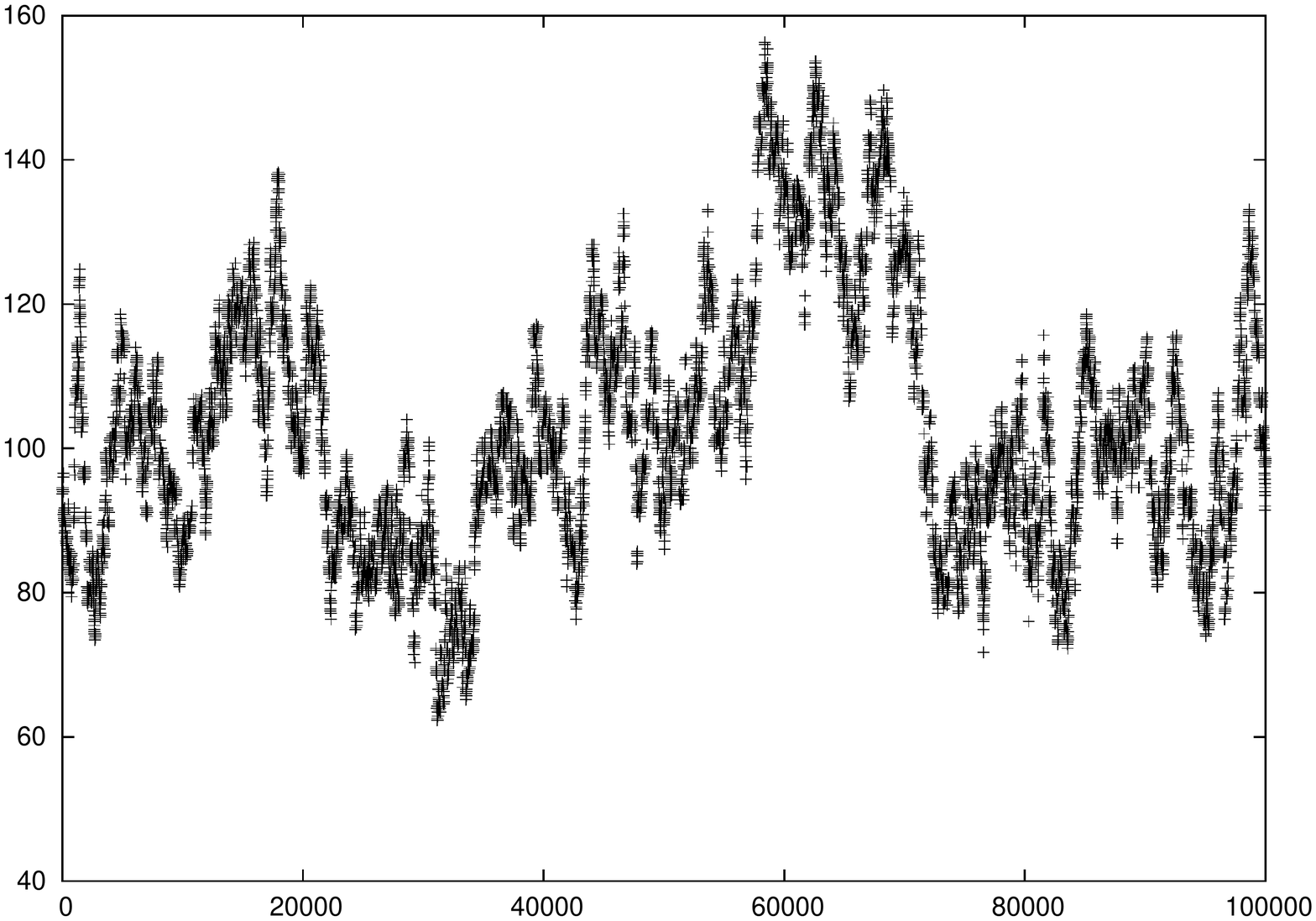}}
\caption{The first $100000$ measurements of the interaction energy (a measurement takes place every $100$ MC steps). Strong autocorrelations are visible in the worm setup.}
\label{en}
\subfigure[Diagonal setup]{\label{errordiag}\includegraphics[width=.49\textwidth]{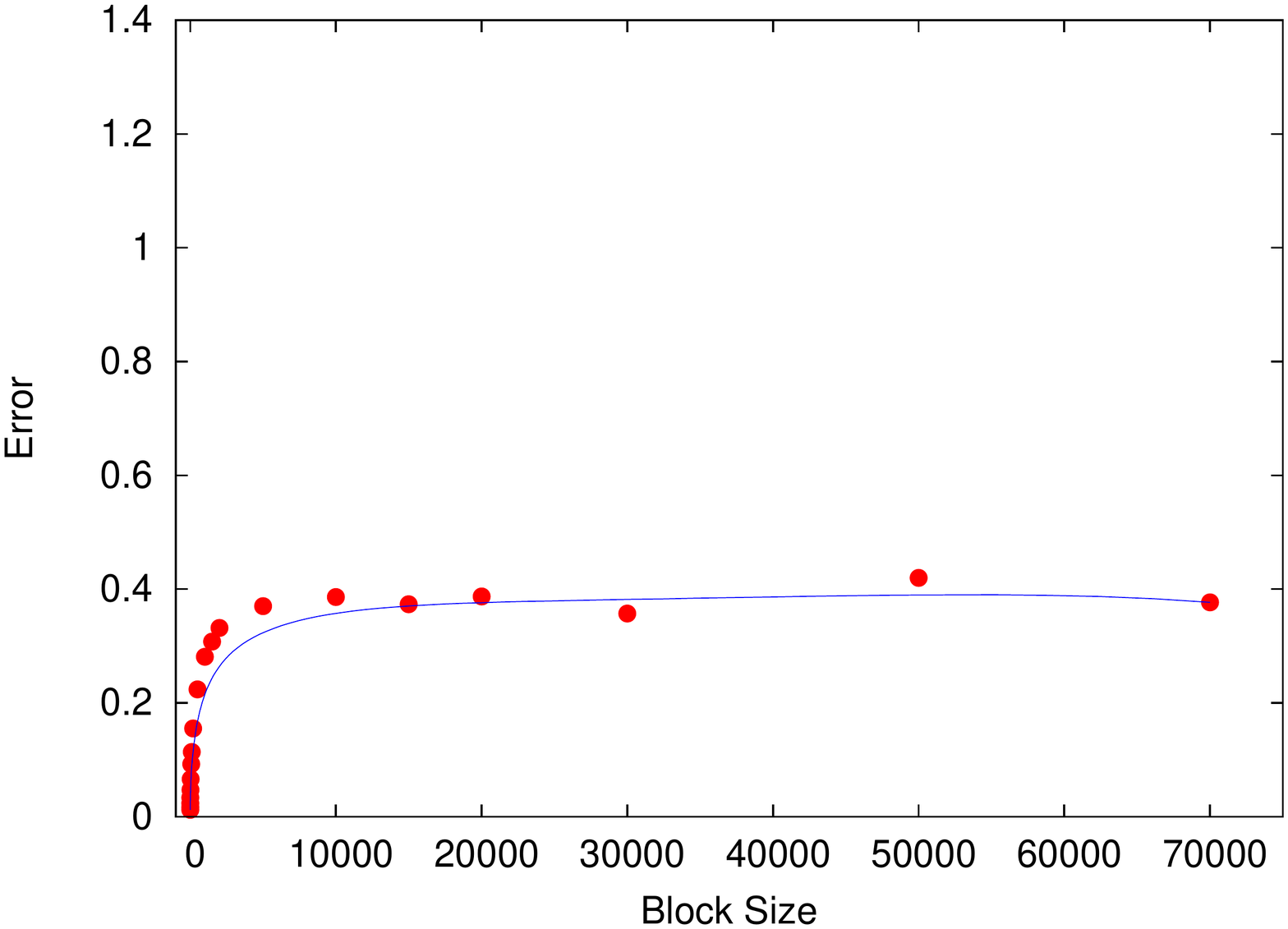}}
\hfill
\subfigure[Worm setup]{\label{errorworm}\includegraphics[width=.49\textwidth]{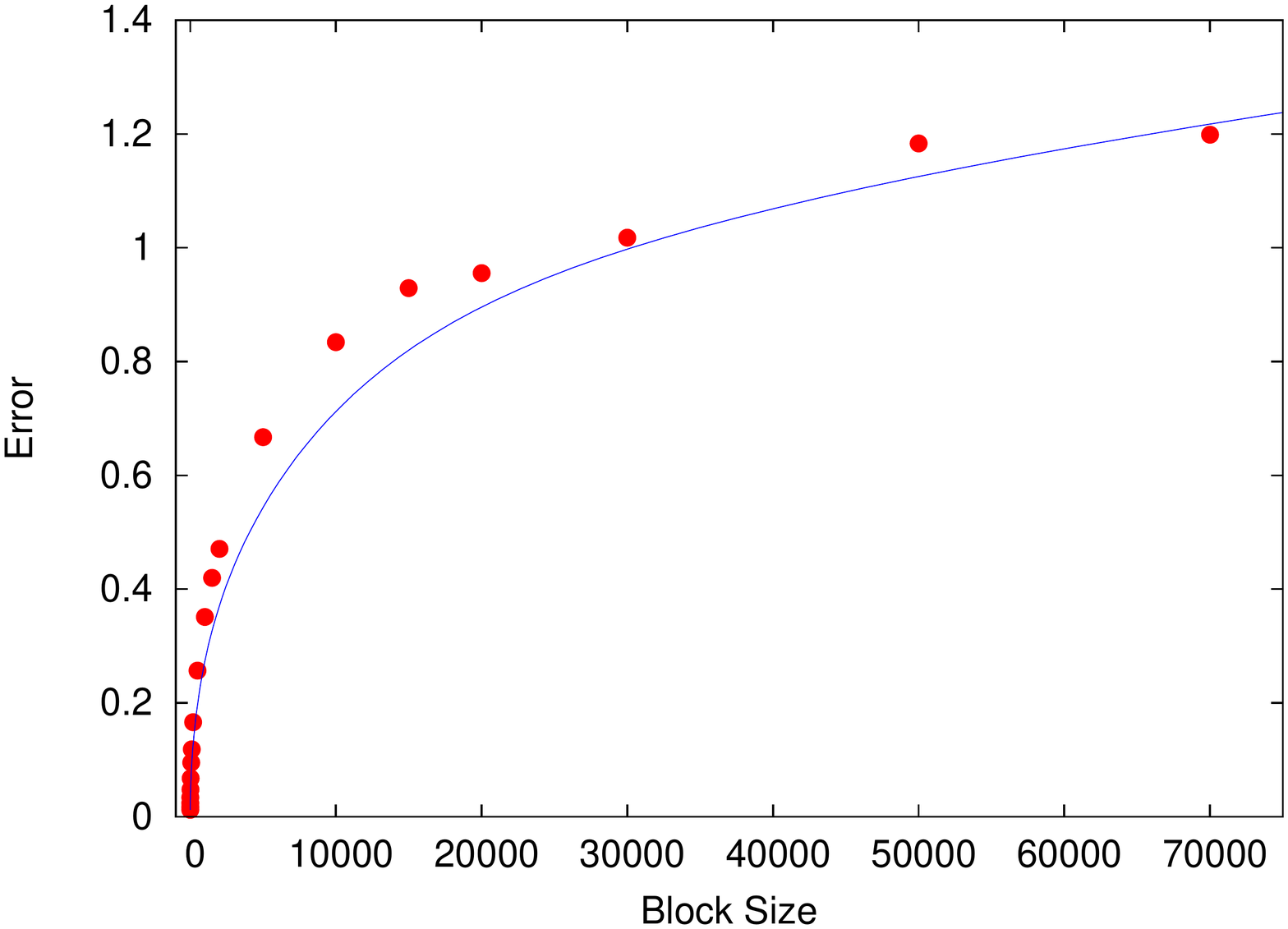}}
\caption{The blocking error analysis of the interaction energy (lines to guide the eye). The blocked error is much higher in the worm setup and continues increasing even for large block sizes.}
\label{err}
\end{center}
\end{figure}

In our study we found that although the worm type addition and removal updates have high acceptance rates, it is at the cost of strong autocorrelations. It is most efficient for the algorithm to successively add and remove the same vertices, so that the configuration does not change significantly, even after many successful updates. To illustrate this compare the measurements of the interaction energy (which is proportional to the diagram order) in the diagonal and the worm setup (figures \ref{en} and \ref{err}). Both simulations used the same parameters and a comparable number of MC steps.

Because of the large measurement error due to autocorrelations the worm setup is effectively less efficient than the standard diagonal setup. These new insights make a modification of the worm algorithm necessary. The goal is hereby to combine the advantages of the diagonal setup (weak autocorrelations) with the ones of the worm setup (high acceptance rates). To achieve this we propose a second type of addition/removal updates:
\begin{itemize}
\item Choose a random 4-point vertex from the configuration (which will act as a worm for this step).
\item Addition: add another 4-point vertex on the same lattice site and in some time interval around the worm.
\item Removal: remove the nearest neighbour of the worm vertex (implies that addition can only be accepted if the new vertex is the nearest neighbour of the worm).
\end{itemize}
This setup still prolongs existing vertex chains, but autocorrelations are significantly reduced since the worm changes with every update. This new type of updates can of course only be employed in addition to the regular diagonal addition and removal updates. It works regardless if the pair of 2-point vertices is present or not in the configuration (the worm addition/removal updates can only take place when the 2-point vertices are present). The acceptance rates for this update are comparable with those for the regular worm updates.

\section{Imbalanced Fermi gas}

The DDMC algorithm presented in the previous sections relies strongly on the assumption of equal densities of the two fermion species. This assumption allows us to write the partition function (\ref{partitionfunction}) as a sum of positive terms only, and consequently to use it as a probability distribution for Monte Carlo sampling. We now present a generalisation of the algorithm to the imbalanced unitary Fermi gas ($\mu_\uparrow\neq\mu_\downarrow$). The imbalanced case is especially interesting for a variety of reasons. A much richer structure of the phase diagram can be observed in this case. The superfluid state has been found to be remarkably stable, however it is also known that at some critical imbalance superfluidity must break down completely \cite{breakdown}. Several experimental studies are already available \cite{expimbalanced}, but to our knowledge numerical studies have so far been only performed at zero temperature \cite{numimbalanced}.

Our goal is to study how an imbalance will affect the critical temperature of the system. In this case a sign problem is present: the function $\rho(S_p)=\frac{1}{Z}(-U)^p\det\mathbf{A}^\uparrow(S_p)\det\mathbf{A}^\downarrow(S_p)$ is no longer positive for all configurations $S_p$ and can thus not be used as a probability distribution. Several methods of dealing with sign problems of this kind are known from lattice QCD, where the introduction of a chemical potential renders the fermionic determinant complex. The most straightforward one is the "phase quenched method", which reduces to a "sign quenched method" in our case. We can write $\rho(S_p)=|\rho(S_p)|\sgn(S_p)$ and use the positive function $|\rho(S_p)|$ as the new probability distribution. For a thermal average this means
\begin{equation}
\langle X\rangle_\rho=\frac{\sum X(S_p)\rho(S_p)}{\sum \rho(S_p)}=\frac{\sum X(S_p)|\rho(S_p)|\sgn(S_p)}{\sum|\rho(S_p)|\sgn(S_p)}=\frac{\langle X\sgn\rangle_{|\rho|}}{\langle\sgn\rangle_{|\rho|}}
\end{equation}
This representation of a thermal average in terms of the new probability distribution $|\rho(S_p)|$ is mathematically equivalent to the usual thermal average. However, numerical errors can become very large if $\langle\sgn\rangle_{|\rho|}\approx0$, as it happens for the expectation value of the phase in QCD. Our studies have shown that for the unitary Fermi gas the sign remains very close to unity for a wide range of imbalances, so that sign quenching is applicable and accurate values for the critical temperature can be obtained. Some preliminary data for the balanced and imbalanced case will be presented in the next section.

\section{Results}

\begin{figure}[b]
\begin{center}
\subfigure[balanced]{\label{balanced52}\includegraphics[width=.49\textwidth]{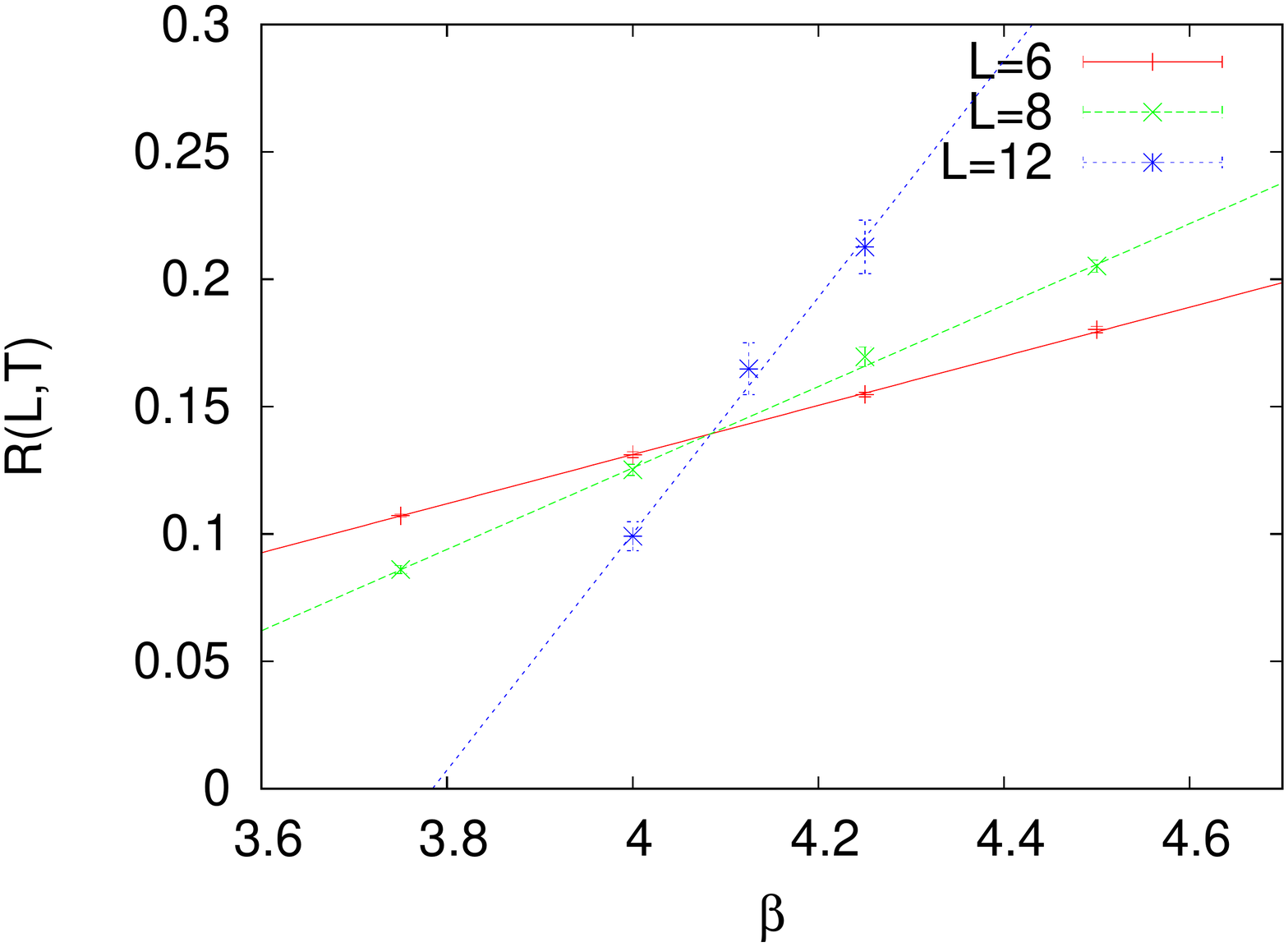}}
\hfill
\subfigure[imbalanced]{\label{imbalanced01}\includegraphics[width=.49\textwidth]{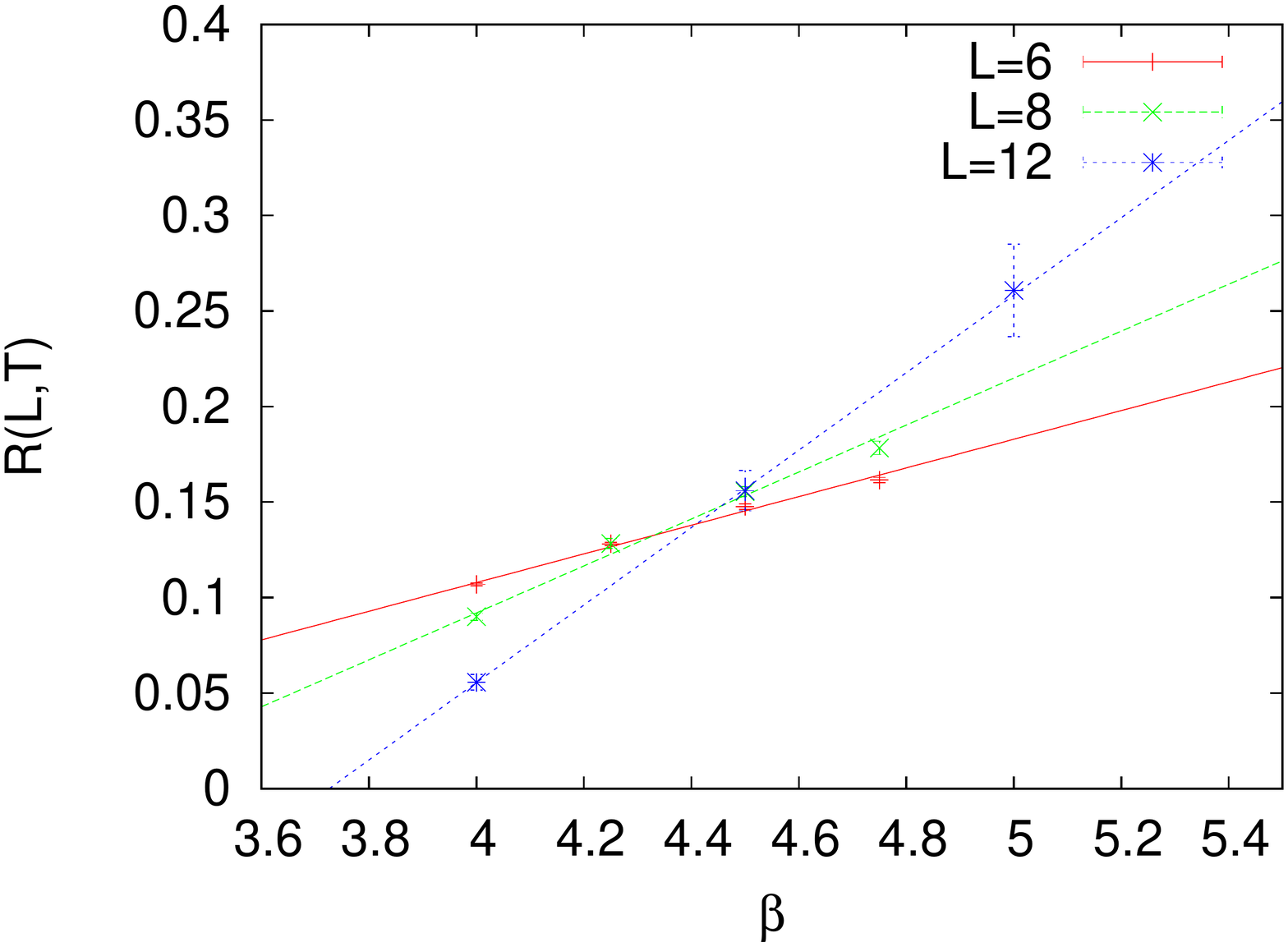}}
\caption{The integrated correlation function $R(L,T)$ is plotted versus the inverse temperature $\beta=1/T$ for different lattice sizes. The lines cross near the critical point, where the order parameter becomes lattice size independent.}
\label{results}
\end{center}
\end{figure}

With the modified worm algorithm we were able to reproduce several values of $T_c$ at different filling factors $\nu$ for the balanced case. Figure \ref{results} shows a typical order parameter analysis. Since the integrated correlation function $R(L,T)$ defined in equation (\ref{intcorrfn}) becomes lattice size independent at the critical point, the $R(L,T)$ curves for different $L$ are expected to cross at $\beta_c=1/T_c$. Due to finite size effects the curves do not cross in exactly one point, but renormalisation group analysis can be applied to extrapolate to the infinite volume limit \cite{main}. For the point in \ref{balanced52} we obtain $\nu^{1/3}=0.542\pm0.003$ and $T_c=(0.087\pm0.002)E_F$, where $E_F$ is the Fermi energy. This agrees well with the results presented in \cite{main}. In figure \ref{imbalanced01} we present the crossing of the order parameter lines for the unitary Fermi gas with an imbalance of $|\Delta\mu|=(0.0398\pm0.0007)E_F$. In this case the expectation value of the sign was found to be between $0.98$ and approximately $1$ for lattice sizes $L=12,8,6$. We can see that the overall errors are sufficiently small. For the point in \ref{imbalanced01} we obtain $\nu^{1/3}=0.512\pm 0.005$ and $T_c=(0.084\pm0.005)E_F$.

Our studies so far indicate that the sign quenched method can be successfully applied to imbalances of up to about $0.3E_F$. For very large imbalances the sign average becomes close to zero, so that reliable results can no longer be obtained. Further measurements for a wider range of densities and imbalances are in progress. 

\setcounter{secnumdepth}{-1}
\section{Acknowledgments}
This work has made use of the resources provided by the Cambridge High Performance Computing Facility. OG is supported by the German Academic Exchange Service (DAAD), the Engineering and Physical Sciences Research Council (EPSRC) and the Cambridge European Trust.

\end{document}